\documentclass[11pt,e-only,final]{amsart}

\dateposted{Preprint posted electronically on December 10, 2003}
\renewcommand{\volinfo}[1]{} 

\copyrightinfo{2003}{M. Cs\H{u}r\"os, B.Li, and A. Milosavljevic}
\PII{}

\usepackage{graphicx}
\usepackage{latexsym,cmmib57}
\usepackage[psamsfonts]{eucal}
\usepackage[in]{fullpage}
\usepackage{chicago}

\newcommand{\condgen}[6]{{#1}#2 #5 #3 #6 #4}
\newcommand{\bbrd}[1]{\mbox{\rm{I}\kern-.1667em{#1}}}
\newcommand{\EXP}{\mathbb{E}}
\newcommand{\PROB}{\mathbb{P}}

\newcommand{\Probcmd}[2]{\condgen{\PROB}{\Bigl\{}{\Bigm|}{\Bigr\}}{#1}{#2}}

\newcommand{\Expcmd}[2]{\condgen{\EXP}{\Bigl[}{\Bigm|}{\Bigr]}{#1}{#2}}

\newcommand{\myfigheight}{.30\textheight}

\newtheorem{theorem}{Theorem}

\newsavebox{\fmbox}

\newcommand{\captionstyle}{\small}

\newcommand{\nclones}{N} 
\newcommand{\clength}{L} 
\newcommand{\nfrags}{F} 
\newcommand{\flength}{\ell} 
\newcommand{\coverage}{c} 
\newcommand{\glength}{G} 

\newcommand{\clone}{B} 

\newcommand{\sigsize}{n} 

\newcommand{\wgscoverage}{w}
\newcommand{\pcoverage}{a}
\newcommand{\roverlap}{\sigma}
\newcommand{\fpooled}{\mu}
\newcommand{\soverlap}{\vartheta}

\newcommand{\coverlap}{\Theta}

\newcommand{\Pa}{P_0}
\newcommand{\Pb}{P_{\fpooled}^{(\sigsize)}}
\newcommand{\Pc}{P_{\fpooled}^{(\infty)}}

\newcommand{\ilen}{\lambda}

\newcommand{\ndfactor}{\beta}
\newcommand{\pnd}{q}

\newcommand{\rvgaps}{G}
\newcommand{\rvreads}{R}
\newcommand{\gengaps}{\mathcal{G}}
\newcommand{\nreads}{r}
\newcommand{\ngaps}{g}
\newcommand{\mreads}{\lambda}

\begin{document}
\title[CAPSS and CAPS-MAP]{Clone-array pooled shotgun mapping and sequencing:\\ 
		design and analysis of experiments}

\author[M. Cs{\H u}r\"os]{Mikl\'os Cs\H{u}r\"os}\address{MC: 
                D\'epartement d'informatique et de recherche op\'erationnelle, 
                Universit\'e de Montr\'eal,                 
                CP 6128 succ. Centre-Ville, 
                Montr\'eal, Qu\'ebec H3C 3J7, Canada. 
                Phone: +1 (514) 343-6111x1655, Fax: +1 (514) 343-5834.}
                \email{csuros@iro.umontreal.ca}
                \urladdr{http://www.iro.umontreal.ca/\textasciitilde{}csuros/}
\author[B. Li]{Bingshan Li}\address{BL:
				Human Genome Sequencing Center, Department of Molecular and Human Genetics,
				Baylor college of Medicine, Houston, Texas, 77030, USA.}
\author[A. Milosavljevic]{
        Aleksandar Milosavljevic}\address{AM:
              Bioinformatics Research Laboratory,
              Program in Structural and Computational Biology and Molecular Biophysics, and
                Human Genome Sequencing Center --- 
                Department of Molecular and Human Genetics,
                Baylor College of Medicine, 
                Houston, Texas 77030, USA.}
                \email{amilosav@bcm.tmc.edu}
                \urladdr{http://www.brl.bcm.tmc.edu/}

\begin{abstract}
This paper studies sequencing and mapping
methods that rely solely on pooling and shotgun sequencing of clones. 
First, we scrutinize and improve 
the recently proposed Clone-Array Pooled 
Shotgun Sequencing (CAPSS) method,
which delivers a BAC-linked assembly of a whole genome sequence.
Secondly, we introduce a novel physical mapping method, called  
{\em Clone-Array Pooled Shotgun Mapping} (CAPS-MAP), which
computes the physical ordering of BACs in a random library.
Both CAPSS and CAPS-MAP  
construct subclone libraries from 
pooled genomic BAC clones.

We propose 
algorithmic and experimental improvements
that make CAPSS a viable option for
sequencing a set of BACs. We provide the first
probabilistic model of CAPSS sequencing progress. The model leads to
theoretical results supporting previous, less formal arguments on 
the practicality of CAPSS. 
We demonstrate the 
usefulness of CAPS-MAP for clone overlap detection
with a probabilistic analysis,
and a simulated assembly
of the Drosophila melanogaster genome.
Our analysis indicates that CAPS-MAP is well-suited for
detecting BAC overlaps in a highly redundant library,
relying on a low amount of shotgun sequence information.
Consequently, it is a practical method 
for computing the physical ordering of clones in 
a random library, without requiring additional clone fingerprinting.
Since CAPS-MAP requires only shotgun sequence reads, 
it can be seamlessly incorporated into a 
sequencing project with almost no experimental overhead. 
\end{abstract}


\maketitle

%

\section{Introduction}
In a {\em hierarchical approach} to large genome sequencing,
one first breaks many genome copies into random fragments.
A {\em library} is constructed by cloning the fragments,
typically as  
{\em Bacterial Artificial
Chromosome} inserts (BACs).
Some BACs in the library are selected for complete sequencing. 
Each selected BAC sequence is assembled individually 
using the shotgun method: a {\em subclone} library 
is prepared by cloning short fragments of the BAC. 
Subsequently, sequence {\em reads}
are produced from a sufficient number of randomly chosen subclones. 
The reads are assembled algorithmically into the BAC sequence.
An alternative to the hierarchical, or {\em clone-by-clone}, 
strategy is the {\em whole-genome shotgun} approach~\shortcite{WGS},
which employs a few (essentially 1--3) subclone libraries prepared 
from the entire genome, 
without resorting to an intermediate BAC library. 
The main advantage of the whole-genome approach 
is that it eliminates 
the need to prepare tens of thousands of subclone libraries to
sequence a mammalian genome. 
However, it is generally an inadequate strategy for finishing the assembly
of such large repeat-rich genomes. 
For a review of contemporary sequencing methodologies, see, e.g., 
\shortciteN{Sequencing.review}.

%
%

A new BAC-based sequencing strategy, called Clone-Array Pooled 
Shotgun Sequencing (CAPSS), was proposed recently~\shortcite{CAPSS}.
CAPSS assembles the complete sequences of individual BACs 
as does the clone-by-clone approach, but requires a much smaller 
number of subclone library preparations.
The strategy is currently being applied for the first time on a 
genome scale in the context of sequencing the honey bee genome. 
This paper provides the theory for the design and analysis of pooling-based 
genome projects. It also introduces the CAPS-MAP method for  
physical mapping,
and transversal pooling designs for both CAPSS and CAPS-MAP,
thereby laying the theoretical foundation for pooling-based genome-scale 
sequencing projects.

\begin{figure}
\centerline{\includegraphics[height=0.3\textheight]{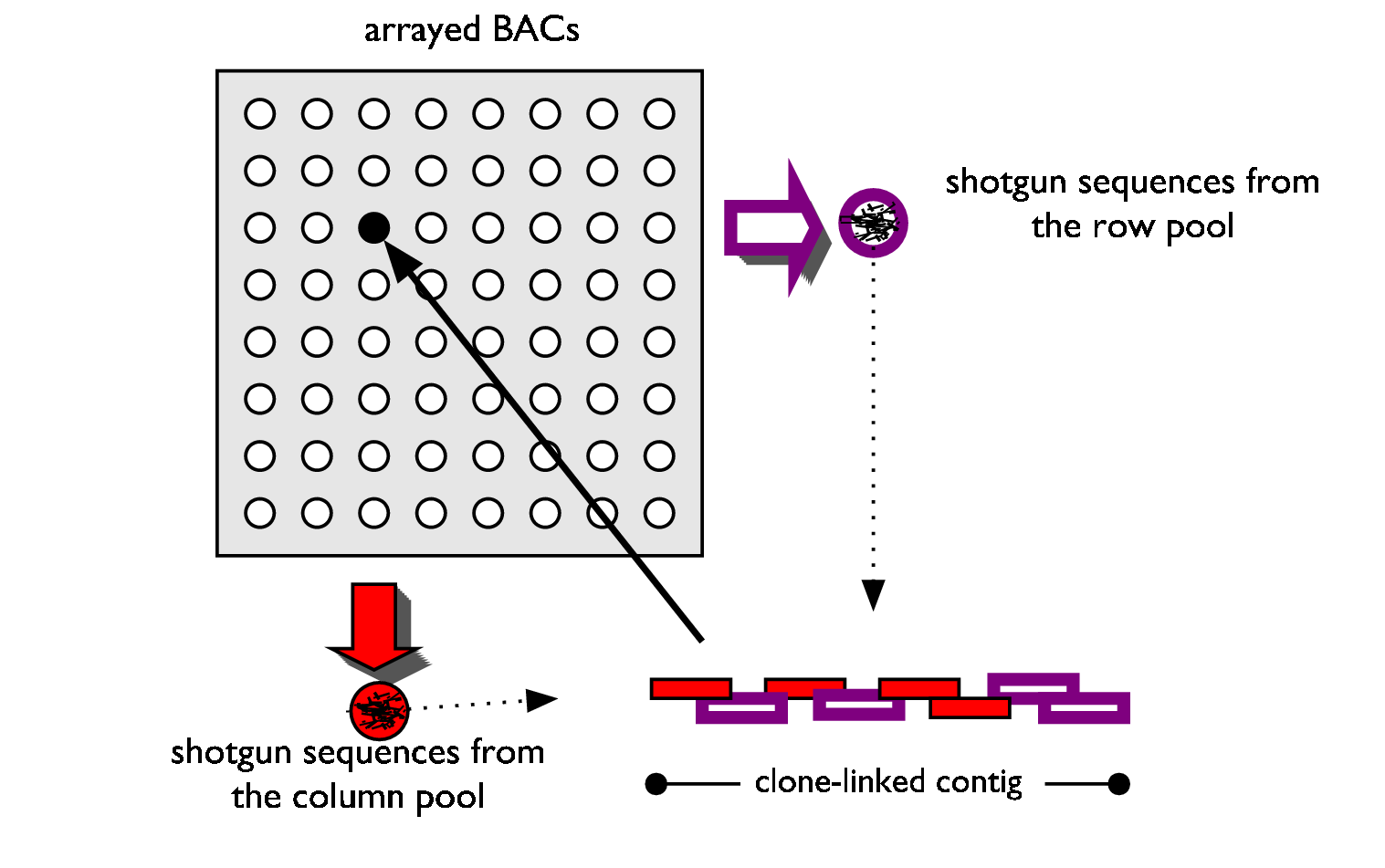}}
\caption[CAPSS]{\captionstyle
CAPSS strategy for arrayed BACs.
DNA extracted from each clone is pooled together with 
other clones in the same row 
and column.  
Subclone libraries are prepared from the pools,
and shotgun sequences are collected from the 
sublibraries. Sequences are assembled into contigs.
If a contig contains sequences from a
row and a column pool's sublibrary,
the contig is assigned to the BAC at the 
intersection of the row and the column.}
\label{fig:capss}
\end{figure}

In a clone-by-clone approach,
BACs are sequenced independently:
one subclone library is constructed 
for every clone. 
In contrast, DNA from BACs are pooled together in a CAPSS approach, 
and subclone libraries are prepared from the pools. 
A CAPSS experiment is designed so 
that the number of subclone libraries is much smaller than the number
of clones, yet the pooling design enables the assembly 
of individual clone sequences. In what follows, 
by {\em pooled shotgun} (CAPS) sequences we mean
shotgun sequence reads collected from a 
subclone library that was constructed using pooled BACs.
For the computational
aspects of sequence assembly, pooled shotgun sequences are random subsequences
originating from a set of clone sequences.

The original CAPSS proposal of \shortciteN{CAPSS} relied 
on a simple rectangular 
design defined by an array layout of BACs 
(Figure~\ref{fig:capss}). The pools correspond to the rows and columns. An array layout
reduces the number of shotgun library preparations to 
the square root of the number of BACs when compared 
to clone-by-clone sequencing. This reduction can be important
in case of a mammalian genome,
for which
even a minimally overlapping tiling path contains between twenty and thirty
thousand clones~\shortcite{human.genome}.

This paper has two goals. First, after pointing out some 
shortcomings of the original CAPSS proposal, we propose 
algorithmic and experimental improvements
that make CAPSS a viable option for
sequencing a set of BACs. 
Specifically, we apply transversal pooling designs to increase the accuracy of CAPSS,  
which we previously developed for the PGI method of comparative 
physical mapping that also uses pooled shotgun sequencing~\shortcite{PGI.conf}. 
We provide the first
probabilistic model of CAPSS sequencing progress.
The model leads to
theoretical results supporting previous, less formal arguments on 
the practicality of CAPSS.

The paper's second goal is to introduce the {\em Clone-Array Pooled Shotgun
Mapping (CAPS-MAP)} method to detect clone overlaps in a random BAC library.
The information on clone overlaps is used to compute the physical 
ordering of clones in 
the library, without requiring additional clone fingerprinting.
CAPS-MAP operates in the same experimental framework as CAPSS.  
It needs only
shotgun sequences, which makes
it a cost-effective method that can be seamlessly integrated into a 
sequencing project with very little experimental overhead. 
We demonstrate the usefulness of CAPS-MAP for clone overlap detection
with a probabilistic analysis.
In addition to the theoretical results, we illustrate the method's performance 
in a simulated project using the Drosophila genome assembly.

\section{Transversal designs}
It was proposed by \shortciteN{CAPSS} that CAPSS be
used in hybrid projects, combining 
whole-genome shotgun (WGS) and pooled shotgun (CAPS) sequences. 
The motivation is that the pooled shotgun sequences 
can provide the localization information 
for the whole-genome shotgun sequences 
so that the latter can be used for a 
clone-linked assembly.
After WGS and CAPS sequences from a set of pools 
are assembled into contigs,
the contigs need to be mapped to individual BACs.
There are a few challenges to contig mapping.
We mention here three 
main problems: false negatives, ambiguities, and false mapping.
A false negative refers to a situation where a BAC is not sampled in a 
pool it is included in, due to the low number of CAPS sequences 
collected.
A false negative for a simple rectangular design means that 
no contigs can be mapped to the BAC. 
Ambiguities and false mappings are caused by overlapping clones, 
or more generally, by clones that have highly similar regions. 
The mapping of a contig is ambiguous if it is not possible to decide
which clones the contig should be assigned to, in cases where 
two or more clone sets are equally likely choices for the mapping.
False mapping occurs when an insufficient number of CAPS
sequences are collected, and a contig 
that covers overlapping BACs gets assigned to the wrong clone or clone set.  
 
One strategy used to overcome the mapping problems
involves transversal pooling designs~\shortcite{PGI.conf,CGT}. 
For a transversal design with~$\sigsize$ pool sets, 
every clone is included in exactly one pool of each pool set, and any subset 
with two of those pools uniquely identifies the clone. 
Half of the pool
sets are designated as column pools, and the other half as row pools 
to realize the design with the help of BAC arrays.
Using a transversal
double-array design (i.e., one with four pool sets), 
the same set of BACs is independently arrayed twice. 
Each of the two resulting 
arrays contains the
same set of BACs.
Thus, each BAC ends up being sampled in two column-pools and
two row-pools. One of the arrays contains an arbitrary arrangement of
BACs, while the other is ``reshuffled'' relative to the first.
More generally, 
clones can be arranged on~$d$ reshuffled arrays using 
a transversal pooling design with~$\sigsize=2d$ pool sets.

The number of arrays in a transversal design may be adjusted to allow
unambiguous and correct contig mapping for any redundancy in a BAC
library. Specifically, it can be shown~\shortcite{PGI.conf,CGT} 
that a $d$-array transversal design can accurately resolve BACs at
up to $(2d-1)$X redundancy.
We previously described and analyzed transversal designs in
the context of pooled shotgun experiments~\shortcite{PGI.conf} 
and compared their performance to
other designs. Even though our analysis was performed for 
the Pooled Genomic Indexing (PGI) method 
in the context of comparative physical mapping, 
the results are generally 
valid for CAPSS and CAPS-MAP as well. Specifically, 
our results indicate that transversal designs 
reduce the frequency of false negatives and false mappings
when compared to a simple rectangular design.
Furthermore, when compared to other more complicated designs, 
they achieve an optimal balance between the number of shotgun 
library preparations and the frequency of contig mapping problems.
Transversal designs also enjoy a practical advantage 
over more complicated combinatorial designs, in that they are 
readily implemented using existing automated clone arraying technologies.

When a transversal design is used, 
contig mapping can be implemented very efficiently,
based on an algorithm that runs in~$O(\nclones+M)$ time for 
mapping~$M$ contigs onto~$\nclones$ BACs. Without going into 
details, 
the main idea is to first build in~$O(\nclones)$ time 
a hash table that maps pool pairs to BACs. Based on the property of transversal designs
that two pools identify a clone, this table contains all pool pairs that identify a 
unique clone. For each contig, it takes~$O(1)$ time using the hash table to 
either identify the most likely clone set to which 
the contig can be mapped, or to declare the contig ambiguous.   

\section{Sequence assembly}\label{sec:capss} 
This section analyzes CAPSS progress in a hybrid 
project that uses whole-genome  
and pooled shotgun sequences. CAPS sequences are collected 
using a transversal design with~$\sigsize$ pool sets, i.e., $\sigsize/2$
arrays. 
In order to derive a probabilistic model for such experiments,
we introduce some standard 
simplifying assumptions and the following notations.
Assume that every clone has the same length~$\clength$ 
(100--200 thousand base pairs in practice),
and that each shotgun sequence has the same length~$\flength$ (e.g., 500 bp).
The WGS and CAPS sequences are 
combined and compared to each other to 
find overlaps between them. 
Overlapping sequences form {\em islands}.
Islands with two or more sequences are {\em contigs}.
An overlap between two shotgun sequences 
is detected if it is at least of length~$\soverlap\flength$
where~$0<\vartheta\le 1$. 
Statistics for islands, and gaps between islands 
are well known~\shortcite{LanderWaterman,WendlWaterston}.
We are interested in statistics for
{\em clone-linked contigs}, those that are assigned
to BACs using the pooling information.

Let~$\pcoverage$ be the coverage by CAPS sequences, i.e., 
if~$\nfrags_{\mathrm{p}}$ CAPS sequences
are collected, then~$\pcoverage=\frac{\nfrags_{\mathrm{p}}\flength}{\nclones\clength}$ 
where~$\nclones$ is the total number of clones. 
Let~$\wgscoverage$ denote the coverage by WGS 
sequences, i.e., if~$\nfrags_{\mathrm{w}}$ WGS 
sequences are collected,
then~$\wgscoverage=\frac{\nfrags_{\mathrm{w}}\flength}{\glength}$ 
where~$\glength$ is the genome length.
Notice that~$\wgscoverage=0$ is possible.
Here we consider the simplest case of 
assembling the sequence of a single 
clone that does not overlap with any other clone.
Such a clone is covered by
a total coverage of~$(\pcoverage+\wgscoverage)$. 
Although we concentrate on sequencing a particular clone, 
the transversal design allows the simultaneous sequencing 
of multiple, possibly overlapping clones 
by combining WGS sequences with CAPS sequences from many (or even all) pools. 
Regions of overlapping clones have higher 
coverage since they are covered by
more CAPS sequences than a single clone.
The sequencing of overlapping regions progresses thus 
faster than what is suggested by the statistics 
for a single clone.
We examine the case of assigning contigs to overlapping BACs
in \S\ref{sec:capsmap}.
Two shotgun sequences from different pools suffice to assign a contig to a single BAC. 
In a practical setting, it may be advantageous to require more 
stringent criteria in order to avoid false mappings. 
Theorem~\ref{tm:capss} can be readily adapted for such 
criteria, albeit resulting in bulkier formulas.

\begin{figure}
\begin{tabular}{cc}\includegraphics[width=0.48\textwidth]{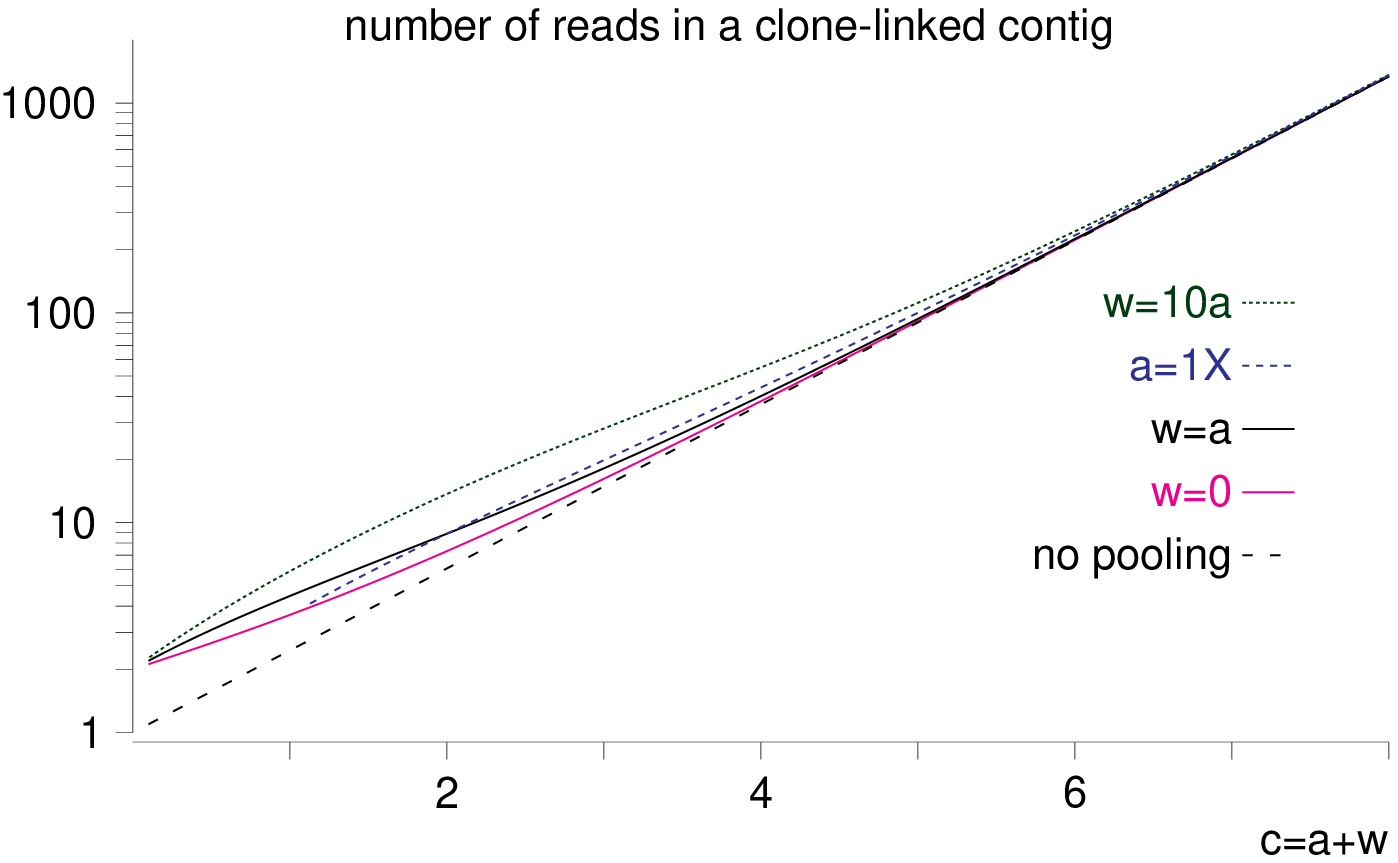} &
\includegraphics[width=0.48\textwidth]{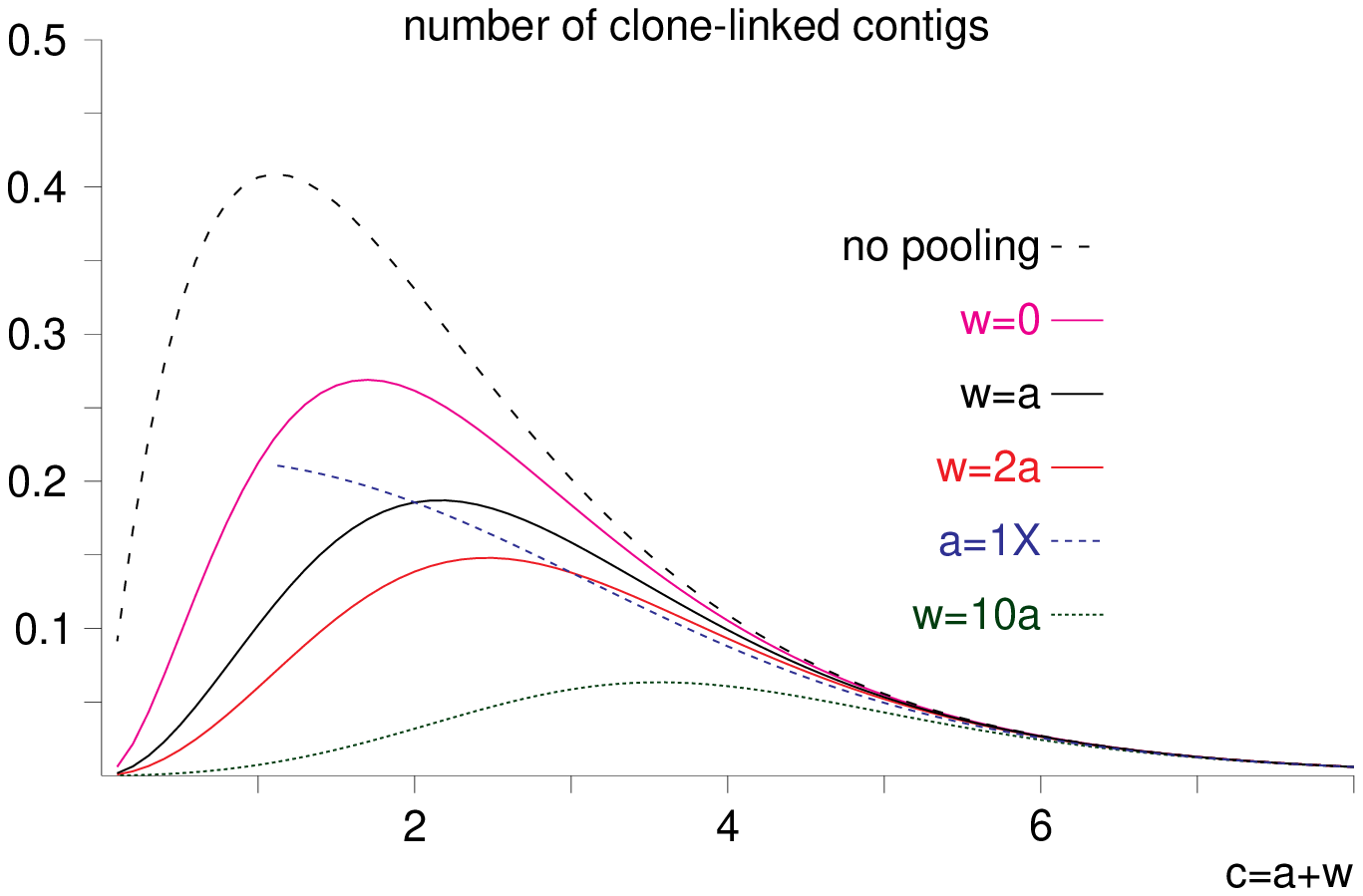}\\
a & b 
\end{tabular}
\caption[CAPSS island statistics]{\captionstyle
CAPSS (Theorem~\ref{tm:capss}): clone-linked contig statistics. The values are calculated 
from Theorem~\ref{tm:capss} for two-array transversal designs
and different pooled coverage levels~$\pcoverage$. Overlaps between shotgun sequences are detected 
with~$\soverlap=0.1$. The number of contigs on the right-hand side is given 
in multiples of~$\clength/\flength$. The abscissa is the total coverage~$\coverage$.}
\label{fig:capss.islands}
\end{figure}

\begin{figure}
\begin{tabular}{cc}\includegraphics[width=0.48\textwidth]{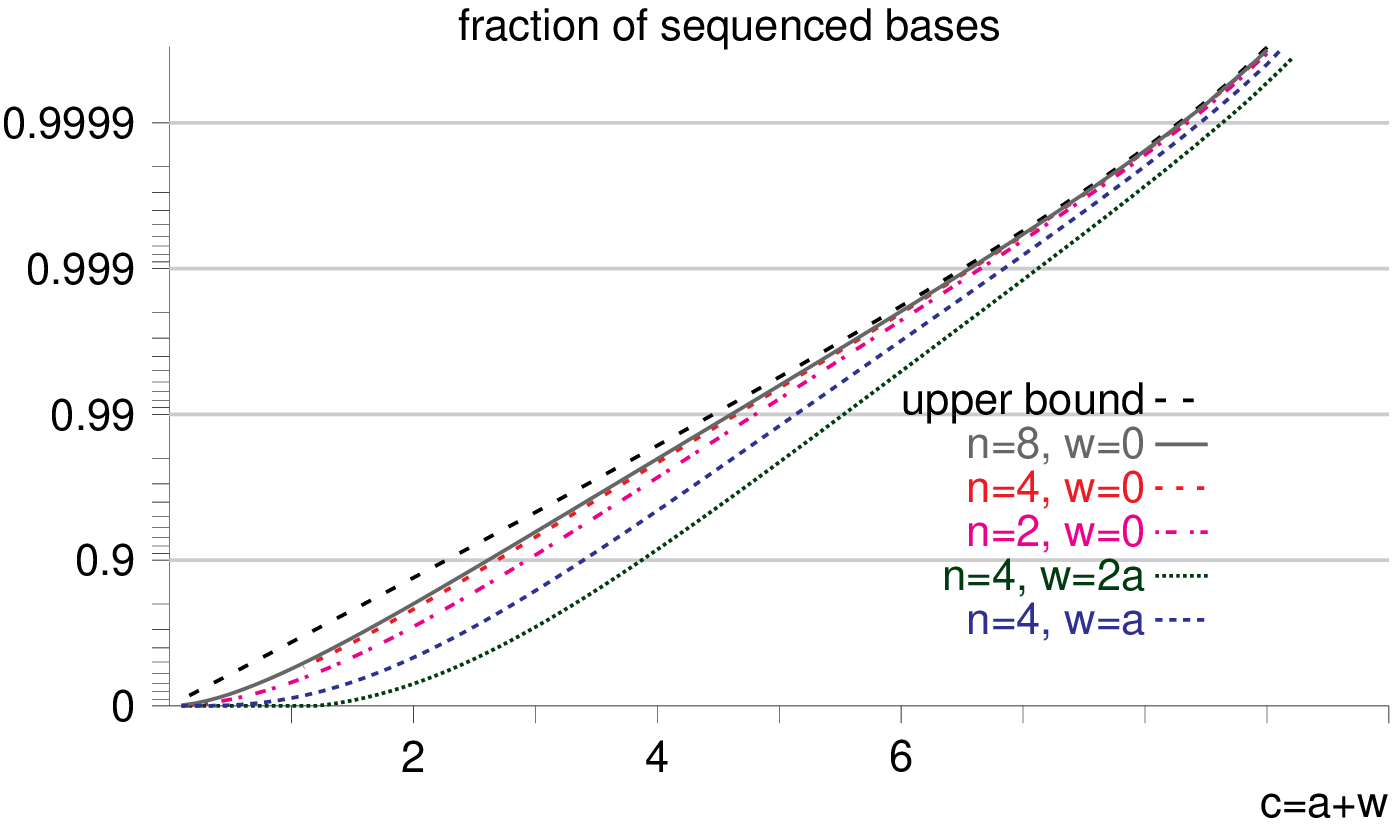} &
\includegraphics[width=0.48\textwidth]{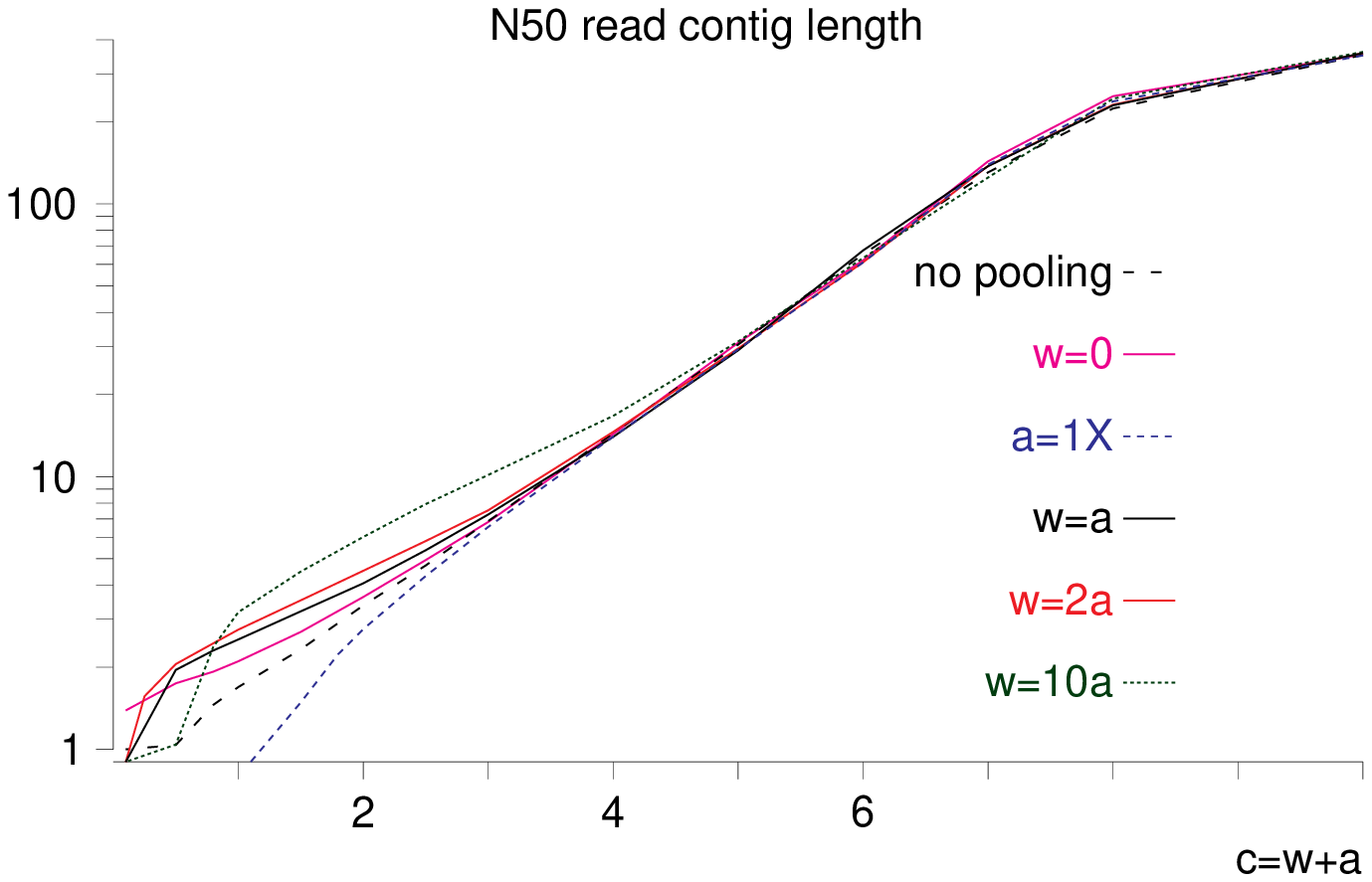}\\
{\captionstyle a} & {\captionstyle b} 
\end{tabular}
\caption[CAPSS sequencing progress]{\captionstyle
	CAPSS (Theorem~\ref{tm:capss}): sequencing progress. 
	The left-hand side plots the 
	fraction of bases covered by clone-linked contigs as
	a function of total coverage ($\coverage=\pcoverage+\wgscoverage$)
	for different designs. Notice that the improvement from two arrays 
	to four arrays ($\sigsize=4$ vs.\ $\sigsize=8$) is marginal.
	The right-hand side plots the N50 values 
	for different designs with two arrays, as multiples of~$\flength$. 
	All values were calculated 
	with shotgun sequence overlap detection~$\soverlap=0.1$. The N50 plot was obtained 
	from simulation: each point is an average of 200 measurements.
}\label{fig:capss.cover}
\end{figure}

Figures~\ref{fig:capss.islands} and~\ref{fig:capss.cover}
compare different experimental designs based on Theorem~\ref{tm:capss}
and simulations. 
Figure~\ref{fig:capss.islands} plots the island statistics from the theorem.
It illustrates that for lower coverages (about $\coverage<4$), the ratio
of pooled shotgun sequences makes a large difference in the sequencing. 
This difference 
is mainly shown in the number of clone-linked contigs, as the 
contig sizes do not differ much. At large coverage levels, when 
sequencing is nearly completed, the impact of pooled sequences is less, 
i.e., WGS sequences can make up for a lower pooled coverage. 
 
Figure~\ref{fig:capss.cover}a 
shows that while more arrays increase the sequencing success, the improvements
are very small after the second array. 
Notice that if the clones are selected from a minimally overlapping tiling path,
then no part of the genome is covered by more than two BACs, 
and thus two arrays suffice for the unambiguous mapping of all contigs 
that cover clone overlaps. 
Figure~\ref{fig:capss.cover}b plots the N50 values. The N50
contig length is the value~$l$ such that half of the 
sequenced nucleotides belong to contigs of length at least~$l$. 
The statistics for all designs converge to
those of a non-pooled sequencing project as the coverage increases. 
In other words, the negative effects of pooling diminish and the  
project progresses just as without pooling:
for example, at total coverage 4--5X, 99\% of the clone is sequenced.

\begin{theorem}\label{tm:capss}
Let~$\roverlap=1-\soverlap$ where~$\soverlap$
is the fraction of length two shotgun sequences must share in order for
the overlap to be detected. 
Consider a BAC that does not overlap with other 
clones.
Define~$\coverage=\wgscoverage+\pcoverage$,
the total coverage. 
Let
$
X_1 = \frac{\wgscoverage+\frac{\pcoverage}{\sigsize}}{\coverage}$, 
$X_2 = \frac{\wgscoverage}{\coverage}$,
and $Y_i=1-(1-e^{-\coverage\roverlap})X_i$ for~$i=1,2$. 

\begin{enumerate}
\item[(i)] 
The expected number of clone-linked contigs covering the clone equals
\begin{equation}\label{eq:num.link}
\frac{\clength}{\flength}\coverage e^{-\coverage\roverlap} p_{\mathrm{link}},
\end{equation}
where
\begin{equation}\label{eq:prob.link}
p_{\mathrm{link}}=
\begin{cases} 
 1-e^{-\coverage\roverlap}
 	\biggl(\sigsize\frac{X_1}{Y_1}-(\sigsize-1)\frac{X_2}{Y_2}\biggr)
			& \text{if $\wgscoverage>0$;}\\*
\frac{1-e^{-\pcoverage\roverlap}}{1+\frac{1}{\sigsize-1}e^{-\pcoverage\roverlap}}
			& \text{if $\wgscoverage=0$.}
\end{cases}
\end{equation}

\item[(ii)] The expected number of shotgun sequences in a clone-linked contig is
\begin{equation}\label{eq:nr.link}
\nfrags_{\mathrm{link}}
= \begin{cases}
\frac{e^{\coverage\roverlap}}{p_{\mathrm{link}}}
\Biggl(1-e^{-2\coverage\roverlap}\biggl(
 				\sigsize\frac{X_1}{Y_1^2}
				-(\sigsize-1)\frac{X_2}{Y_2^2}\biggr)\Biggr) & \text{if $\wgscoverage>0$;}\\
e^{\pcoverage\roverlap}+\frac{1+\frac{1}{\sigsize-1}}{1+\frac{e^{-\pcoverage\roverlap}}{\sigsize-1}}
					& \text{if $\wgscoverage=0$.}
\end{cases}
\end{equation}

\item[(iii)]
Define 
\begin{align}
\label{eq:mk.nolink}
\nfrags_{\mathrm{nolink}} 
=	\frac{\sigsize\frac{X_1}{Y_1^2}-(\sigsize-1)\frac{X_2}{Y_2^2}}{\sigsize\frac{X_1}{Y_1}-(\sigsize-1)\frac{X_2}{Y_2}},
\intertext{and}
\lambda_{\mathrm{CBC}}
=
	\frac{e^{\coverage\roverlap}-1}{\coverage}+\soverlap.
\end{align}
The expected length of a clone-linked contig
can be written as $\flength\ilen_{\mathrm{link}}$
where $\ilen_{\mathrm{link}}$
is bounded as
\begin{equation}\label{eq:len.link}
\frac{\ilen_{\mathrm{CBC}}-\Bigl(\nfrags_{\mathrm{nolink}}\roverlap+\soverlap\Bigr)(1-p_{\mathrm{link}})}{
	p_{\mathrm{link}}}
\le 
	\ilen_{\mathrm{link}}
\le 
	\frac{\ilen_{\mathrm{CBC}}}{p_{\mathrm{link}}}. 
\end{equation}
Furthermore, when~$\fpooled=\pcoverage/\coverage$ is kept constant, 
$\nfrags_{\mathrm{nolink}}$ increases monotonically
with~$\coverage$ and
\begin{equation}\label{eq:len.limit}
\lim_{\coverage\to\infty}\nfrags_{\mathrm{nolink}}
=
\begin{cases}
\fpooled^{-1}
	\frac{(3\sigsize^2-3\sigsize+1)-\fpooled(2\sigsize^2-3\sigsize+1)}{
		(2\sigsize^2-3\sigsize+1)-\fpooled(\sigsize^2-2\sigsize+1)}
		& \text{if $\wgscoverage>0$;}\\
\frac{\sigsize}{\sigsize-1} & \text{if $\wgscoverage=0$.}
\end{cases}
\end{equation}
\end{enumerate}
\end{theorem}

\begin{proof}
The proof relies on a Poisson process model, following the technique
of \shortciteN{Waterman}.   
We model the location of the shotgun sequences as a Poisson process 
with rate~$\coverage$. 
Define~$\fpooled=\pcoverage/\coverage$, the fraction of CAPS sequences. 
Every sequence is either a 
WGS sequence with probability~$(1-\fpooled)$, or
comes from each one of the clone's pools with 
probability~$\fpooled/\sigsize$. 
First we state the well-known facts~\shortcite{LanderWaterman,Waterman} 
about apparent islands, whether or not they are linked to a clone.
The event~$E$ that a given shotgun sequence is the right-hand end of an apparent island 
has probability~$J=\PROB E=e^{-\coverage\roverlap}$. 
For the $k$-th read, define~$M_k$ as the number of reads 
from its right-hand end until the first gap towards the left. 
The probability that an island has~$j$ sequences in it equals
\[
\Probcmd{M_k=j}{E}=(1-J)^{j-1}J.
\]

An island can be mapped to a clone if it contains sequences from at least two pools. 
The probability of mapping the island ending at the $k$-th read 
(event $D_k$)
depends on the number of shotgun sequences in the island. Using inclusion-exclusion:
\begin{multline}\label{eq:dme}
\Probcmd{D_k}{M_k=j}  \\*
\begin{aligned}
& =  1-\sum_{\text{pools}}\Probcmd{\text{CAPS reads from only one pool+WGS}}{M_k=j}\\
& +(\sigsize-1) \Probcmd{\text{only WGS reads}}{M_k=j}\\*
& =  1-\sigsize\Bigl(1-\frac{\sigsize-1}{\sigsize}\fpooled\Bigr)^{j}+(\sigsize-1)(1-\fpooled)^j.
\end{aligned}
\end{multline}

By Equation~\eqref{eq:dme}, 
the number of shotgun sequences in a clone-linked island is distributed 
by the probabilities
\begin{multline}\label{eq:prob.dm}
\Probcmd{D_k,M_k=j}{E}  = \Probcmd{D_k}{M_k=j, E}\Probcmd{M_k=j}{E} \\*
\begin{aligned}
& =  \biggl(1-\sigsize\Bigl(1-\frac{\sigsize-1}{\sigsize}\fpooled\Bigr)^{j}
					+(\sigsize-1)(1-\fpooled)^j\biggr)
				(1-J)^{j-1}J\\
& = \Pa(j) - \sigsize\Pb(j) + (\sigsize-1)\Pc(j).
\end{aligned}
\end{multline}
with
\begin{subequations}\label{eq:px}
\begin{align}
\Pa(j) & =  (1-J)^{j-1}J; \\*
\Pb(j) & = 	\biggl((1-J)\Bigl(1-\frac{\sigsize-1}{\sigsize}\fpooled\Bigr)
				\biggr)^{j-1} 
					J\Bigl(1-\frac{\sigsize-1}{\sigsize}\fpooled\Bigr);\\*
\Pc(j) & = \Bigl((1-J)(1-\fpooled)\Bigr)^{j-1} J(1-\fpooled).
\end{align}
\end{subequations}

Now, for all~$0< z\le 1$, 
\begin{equation}
\sum_{j=1}^\infty (1-z)^{j-1} = \frac{1}{z}; \quad
\sum_{j=1}^\infty j(1-z)^{j-1} = \frac{1}{z^2}. \label{eq:z}
\end{equation}
Using Equation~\eqref{eq:z},
\begin{align*}
\Probcmd{D_k}{E} 
& =\sum_{j=1}^\infty \Probcmd{D_k, M_k=j}{E} \\*
& = 1-\frac{\sigsize J(1-\frac{\sigsize-1}{\sigsize}\fpooled)}{1-(1-J)\Bigl(1-\frac{\sigsize-1}{\sigsize}\fpooled\Bigr)}
	+\frac{(\sigsize-1)J(1-\fpooled)}{1-(1-J)(1-\fpooled)}.
\end{align*}
In Equation~\eqref{eq:prob.link},
$p_{\mathrm{link}}=\Probcmd{D_k}{E}$.
Equation~\eqref{eq:num.link} follows from 
the fact that the expected number of shotgun fragments covering the clone
equals~$\coverage\clength/\flength$.

By definition of the conditional probability,
\[
\Probcmd{M_k=j}{D_k, E}=
\frac{\Probcmd{D_k,M_k=j}{E}}{\Probcmd{D_k}{E}} = \frac{\Pa(j)-\sigsize\Pb(j)+(\sigsize-1)\Pc(j)}{p_{\mathrm{link}}},
\]
where the values can be plugged in from 
Equations~\eqref{eq:prob.link} and~\eqref{eq:px}.
By Equation~\eqref{eq:z},
\begin{equation}\label{eq:mk}
\Expcmd{M_k}{D_k,E} 
= \frac{p_{\mathrm{link}}^{-1}}{J}
	\Biggl(1-\frac{nJ^2\Bigl(1-\frac{\sigsize}{\sigsize-1}\fpooled\Bigr)}{
				\Bigl(1-(1-J)(1-\frac{\sigsize-1}{\sigsize}\fpooled)\Bigr)^{2}}
			+\frac{(\sigsize-1)J^2(1-\fpooled)}{
				\Bigl(1-(1-J)(1-\fpooled)\Bigr)^2}
	\Biggr),
\end{equation}
which corresponds to~(ii)
with~$\nfrags_{\mathrm{link}}=\Expcmd{M_k}{D_k,E}$.
It is interesting to notice that
when~$\fpooled=1$, in Equation~\eqref{eq:mk},
\[
\frac2{J(\roverlap)} \ge \Expcmd{M_k}{D_k,E} > \frac1{J(\roverlap)},
\]
and that $\Expcmd{M_k}{D_k,E}J^{-1}(\roverlap)$ decreases when the coverage~$\coverage$ increases.

By Equation~\eqref{eq:prob.link},
\begin{equation}\label{eq:nolink}
\Probcmd{\overline{D_k}}{E} 
= 1-p_{\mathrm{link}}
=1-J\frac{1-\Bigl(1-J\Bigr)\Bigl(1-\frac{\sigsize-1}{\sigsize}\fpooled\Bigr)(1-\fpooled)}{%
	\biggl(1-\Bigl(1-J\Bigr)\Bigl(1-\frac{\sigsize-1}{\sigsize}\fpooled\Bigr)\biggr)
	\biggl(1-\Bigl(1-J\Bigr)\Bigl(1-\fpooled\Bigr)\biggr)}.
\end{equation}
The expected number of shotgun sequences in an island that is not mapped to a clone equals
\[
\Expcmd{M_k}{\overline{D_k},E}
=\frac{\Expcmd{M_k}{E}-\Expcmd{M_k}{D_k,E}\Probcmd{D_k}{E}}{\Probcmd{\overline{D_k}}{E}}.
\]
Using~$\Expcmd{M_k}{E}=J^{-1}$ and Equations~\eqref{eq:mk}, \eqref{eq:prob.link}, 
and~\eqref{eq:nolink},
we get Equation~\eqref{eq:mk.nolink} with the notation
$\nfrags_{\mathrm{nolink}}=\Expcmd{M_k}{\overline{D_k},E}$.

Let~$\flength\ilen_k$ be the length of the island ending with the $k$-th sequence.
The length of a non-linked island can be bounded as
$\flength\Expcmd{\ilen_{\mathrm{nolink}}}{\overline{D_k},E}$
with 
\[
1\le 
\Expcmd{\ilen_{\mathrm{nolink}}}{\overline{D_k},E}
\le \Expcmd{M_k}{\overline{D_k},E}\roverlap+\soverlap.
\]
The bounds of Equation~\eqref{eq:len.link} follow from
\[
\Expcmd{\ilen_k}{D_k,E}
 =  \frac{\Expcmd{\ilen_k}{E}-\Expcmd{\ilen_k}{\overline{D_k},E}\Probcmd{\overline{D_k}}{E}}{
	\Probcmd{\overline{D_k}}{E}},
\]
where $\Expcmd{\ilen_k}{E}=\ilen_{\mathrm{CBC}}=\frac{J^{-1}-1}{\coverage}+\soverlap$
\shortcite{Waterman}.
\end{proof}

The value~$\ilen_{\mathrm{CBC}}$ in the 
theorem is the expected island length in a non-pooled sequencing project. 
By Equation~\eqref{eq:len.limit}, and
the fact that $\lim_{\coverage\to\infty}p_{\mathrm{link}}=1$,
we have
$\lim_{\coverage\to\infty}\ilen_{\mathrm{link}}=\ilen_{\mathrm{CBC}}$
when the ratio of CAPS sequences is kept constant. 
This limit result is not surprising given that
every island can be assigned to a clone with near certainty when the
sequence read coverage is large.
 
\section{Clone overlap detection}\label{sec:capsmap}
\begin{figure}
\centerline{\includegraphics[height=0\myfigheight]{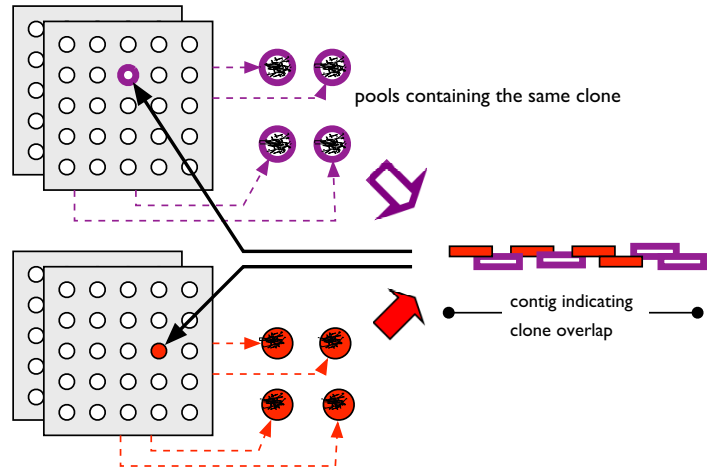}}
\caption[CAPS-MAP]{\captionstyle
CAPS-MAP detects overlaps between clones by identifying situations where
a read contig maps simultaneously to two clones. This figure illustrates a
transversal pooling design with two clone groups and two arrays per group.
The transversal design guarantees that the intersection of any
two pools out of four possible for each BAC (two row and two
column pools) uniquely identifies the BAC. 
Note that overlaps between clones on the same array can also be 
detected by a transversal design.}\label{fig:capsm} 
\end{figure}

\begin{figure}
\centerline{\includegraphics[height=\myfigheight]{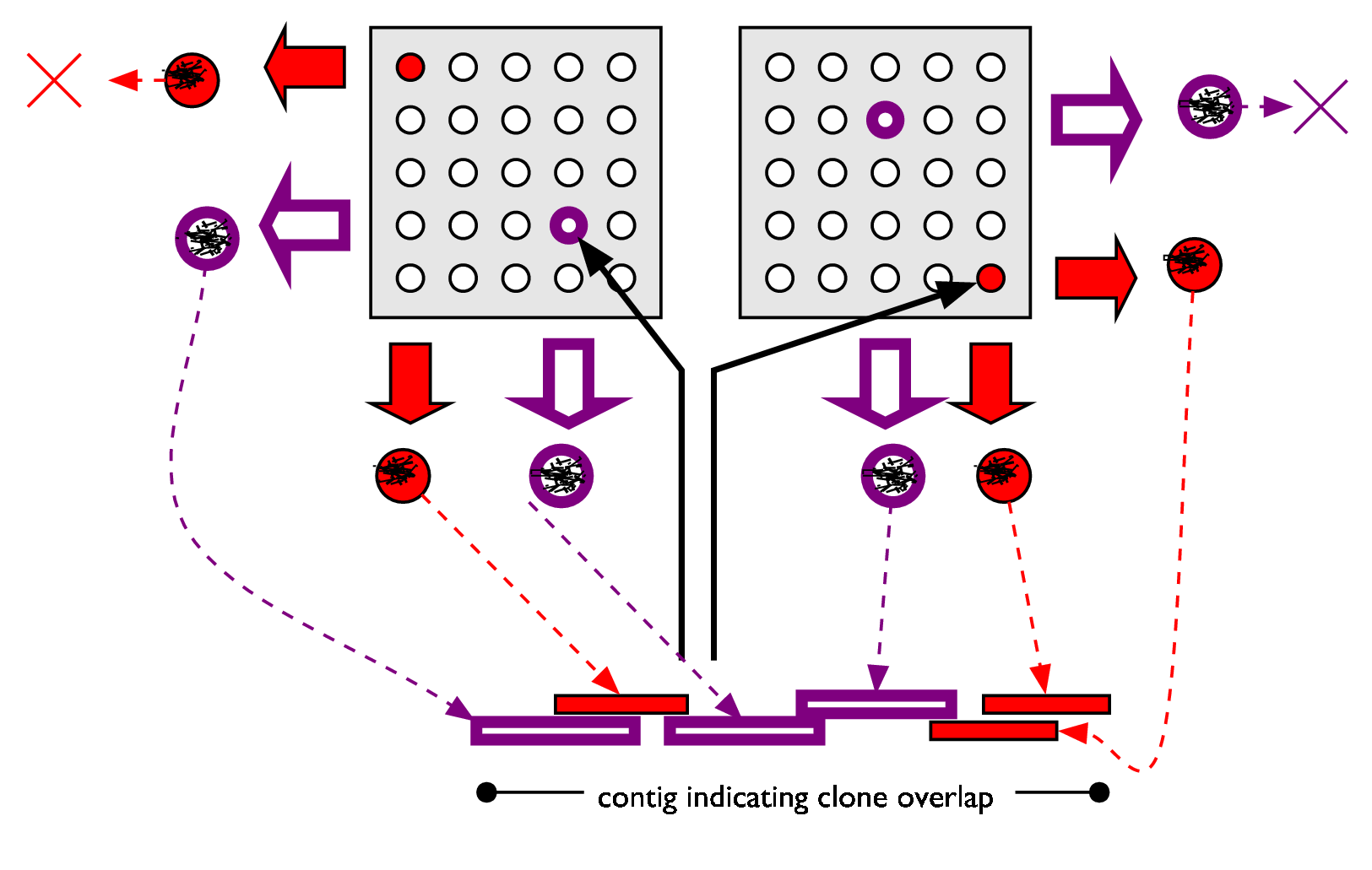}}
\caption{
Overlaps between clones on the same array can also be 
detected by a transversal design, even in the presence of false negatives,
i.e., situations where a particular BAC
is not represented in a particular pool. Specifically, overlap between
the two BACs illustrated in the figure is detected despite the fact that
each BAC is sampled in only three pools.}\label{fig:capsm.fn}
\end{figure}

The key observation for this section is that a transversal design
makes it possible to map a contig unambiguously to more than one 
BAC at once. Now, a contig that is mapped to two clones simultaneously can be 
viewed as evidence that the two clones overlap. Taking the idea further,
an entire set of BACs can be tested for overlaps in this manner,
which leads us to the Clone-Array Pooled Shotgun Mapping (CAPS-MAP) method
that is
described as follows. A redundant collection of random BACs covering a large
genome is grouped into subsets of size~$q^2$. 
Pooled shotgun sequence reads are collected from each clone group using a 
transversal design with~$d$ arrays of size~$q\times q$.
Partitioning into subsets may be dictated by the practical 
concerns of chemistry, biology and robotic automation.
For array sizes that are multiples of 8 or 12 or both 
(yielding standard dimensions of a 
96-well microtiter plate), such as~$q=24$, or~$q=48$, 
there exist known~\shortcite{design.handbook} transversal designs.
A pooling design with a few ($d=2,3,4$) arrays suffice to compute the
physical ordering of BACs in the library, depending on the 
library's redundancy and the array sizes. 
In addition to the CAPS sequences, 
WGS sequences are used to increase read
contig lengths. 
The shotgun sequences are compared to each other to 
find the overlaps between them, 
and are assembled into contigs. Contigs that map
unambiguously to more than one clone are taken as evidence that the clones overlap.
See Figures~\ref{fig:capsm} and~\ref{fig:capsm.fn} for illustrations. 
The clone overlap information can then be used to
compute the physical ordering of the BACs in the library, and 
to select a minimal tiling path for complete sequencing, 
just as if the overlaps were detected using a
fingerprinting scheme \shortcite{map.sequenceready}.

Theorem~\ref{tm:capsm} considers the case of detecting an overlap between
two clones in different clone groups. Similar analyses can be carried out 
for more general cases with more overlapping clones, or clones 
in the same clone group, resulting in more cumbersome formulas. 

\begin{figure}
\centerline{
\includegraphics[height=0\myfigheight]{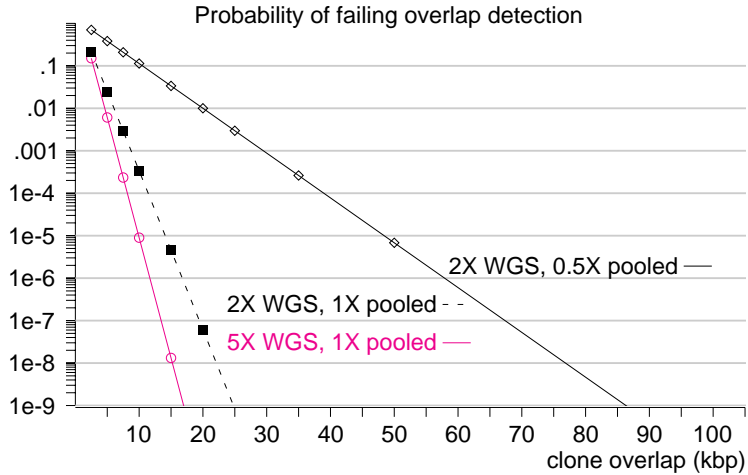}}
\caption[CAPS-MAP statistics]{\captionstyle
	Clone overlap detection.
	The graph shows the probability of not detecting an overlap between
	two clones, as a function of the overlap size. The plots were calculated by a Monte-Carlo 
	method using Theorem~\ref{tm:capsm}.
	All plots use~$\soverlap=0.1$ for shotgun sequence overlap detection, and
	$\flength=500$ for shotgun sequence length.
}
	\label{fig:capsm.probs}
\end{figure}

Figure~\ref{fig:capsm.probs} plots
the overlap detection probabilities 
in a few scenarios with different 
amounts of CAPS and WGS sequences.
Based on the figure, the probability of detecting an overlap
increases exponentially toward~1 with the overlap length. The same exponential 
behavior is characteristic of clone anchoring methods for overlap detection
\shortcite{map.anchor}. Consequently, clone contig statistics for CAPS-MAP can be 
calculated using a clone anchoring model with an appropriate 
anchoring process intensity. Clone contig statistics can also be estimated
using a fingerprinting model \shortcite{LanderWaterman} 
by noticing that clone overlaps above a certain length are detected with near certainty.
Figure~\ref{fig:capsm.probs} indicates that using 1X CAPS coverage and 
2--5X WGS coverage, BAC overlaps of more than 20000 bp are detected almost certainly. 
While CAPS-MAP uses only the fact that a contig is mapped to multiple BACs, and 
not the actual contig sequence, the sequence information is used in 
the ensuing sequencing phase, and thus CAPS-MAP represents very little overhead in a 
genome sequencing project.

It is worth pointing out here that CAPS-MAP detects very short, or even 
{\em negative} clone overlaps with non-negligible probability. 
A short region of the genome 
that is not covered by BACs in the library can be bridged by WGS sequences.
The bridging WGS sequences may form a contig with CAPS sequences from the two BACs 
at the gap's ends that can be mapped to the two clones simultaneously. 
This unique feature of CAPS-MAP among clone overlap detection methods 
does not interfere with the calculation of the physical ordering of BACs.  
At the same time, it does decrease the necessary BAC library size for 
sequencing the genome completely.
After the clones are selected for complete sequencing, the
already collected WGS sequences are included in the genome sequence assembly.
Consequently, negative overlaps detected by CAPS-MAP are already covered by 
shotgun sequences
in the sequencing phase, and
pose no additional requirements for shotgun sequence collection.

\begin{theorem}\label{tm:capsm}
Let two clones from different clone groups 
share an overlap. 
Define~$\coverage_2=2\pcoverage+\wgscoverage$,
the total shotgun sequence coverage for the overlap.
Define
\begin{gather*}
\begin{aligned}
\ndfactor_1 & = \frac{\wgscoverage+(1+\frac{1}{\sigsize})\pcoverage}{\coverage_2} & 
\ndfactor_2 & = \frac{\wgscoverage+\pcoverage}{\coverage_2} & 
\ndfactor_3 & = \frac{\wgscoverage+\frac{2\pcoverage}{\sigsize}}{\coverage_2} &
\ndfactor_4 & = \frac{\wgscoverage+\frac{\pcoverage}{\sigsize}}{\coverage_2} &
\ndfactor_5 & = \frac{\wgscoverage}{\coverage_2};
\end{aligned}\\*
\gamma_i=1-(1-e^{-\coverage_2\roverlap})\ndfactor_i
\quad\text{ for $i=1,\dotsc,5$}.
\end{gather*}

\begin{enumerate}
\item[(i)] An apparent island in the 
overlap consisting of~$j>0$ shotgun sequences 
is mapped to the two clones simultaneously 
with probability~$1-\pnd(j)$ where
\begin{equation}\label{eq:pnd}
\pnd(j) 
= 2\sigsize \ndfactor_1^j
	- 2(\sigsize-1) \ndfactor_2^j
	- \sigsize^2 \ndfactor_3^j
	+ 2\sigsize(\sigsize-1) \ndfactor_4^j
	- (\sigsize-1)^2 \ndfactor_5^j
	< 2\sigsize\ndfactor_1^j.
\end{equation}

\item[(ii)] An apparent island covering the overlap is mapped
to the two clones simultaneously with probability
\begin{equation}\label{eq:prob.map}
p_2
= 1-e^{-\coverage_2\roverlap}\biggl(
	2\sigsize\frac{\ndfactor_1}{\gamma_1}
	-2(\sigsize-1)\frac{\ndfactor_2}{\gamma_2}
	-\sigsize^2\frac{\ndfactor_3}{\gamma_3}
	+2\sigsize(\sigsize-1)\frac{\ndfactor_4}{\gamma_4}
	-(\sigsize-1)^2\frac{\ndfactor_5}{\gamma_5}\biggr).
\end{equation}

\end{enumerate}
\end{theorem}

\begin{proof}
The overlap is detected if 
it is covered by an island that can be simultaneously mapped 
to the two clones.
We model the location of the shotgun sequences as a Poisson process 
with rate~$\coverage_2$. 
Define~$\fpooled_2=\frac{2\pcoverage}{\coverage_2}$,
the fraction of CAPS sequences covering the overlap.
Every shotgun sequence is either a 
WGS sequence with probability~$(1-\fpooled_2)$, or
comes from each one of the two clones' pools with 
probability~$\fpooled_2/(2\sigsize)$. 
The event~$E_2$ that a given shotgun sequence is the right-hand end of an apparent island 
has probability~$J_2=\PROB E_2=e^{-\coverage_2\roverlap}$. 
For the $k$-th sequence, define~$M_k$ as the number of sequences 
from its right-hand end until the first gap towards the left. 
The probability that an island has~$j$ sequences in it equals
\[
\Probcmd{M_k=j}{E_2}=\Bigl(1-J_2\Bigr)^{j-1}J_2.
\]
The probability of mapping the island that ends at the $k$-th shotgun sequence
(event $D_k$)
depends on the number of sequences in the island. We 
calculate the probability of event~$\overline{D_k}$ in separate cases. 
Let~$p_{0,0}(j)$ denote the event that the island 
consists of WGS sequence reads only given that it has~$j$ reads. 
Then
\begin{subequations}
\begin{equation}\label{eq:p00}
p_{0,0}(j) = (1-\fpooled_2)^j.
\end{equation}
Let~$p_{0,*}(j)$ denote the event that the island 
consists of CAPS sequences for one clone only and WGS sequences, given that it has~$j$ 
shotgun sequences in it:
\begin{equation}\label{eq:p0x}
p_{0,*}(j)=\Bigl(1-\frac{\fpooled_2}2\Bigr)^j.
\end{equation}
Let~$p_{1,0}(j)$ denote the event that the island 
consists of CAPS sequences from a fixed pool and WGS sequences, given that it has~$j$ 
shotgun sequences in it:
\begin{equation}\label{eq:p10}
p_{1,0}(j)=\Bigl(1-\frac{\sigsize-\frac12}{\sigsize}\fpooled_2\Bigr)^j-p_{0,0}(j).
\end{equation}
Let~$p_{1,1}(j)$ denote the event that the island 
consists of CAPS sequences from a fixed pool for one clone, 
from another fixed pool for the other clone,
and WGS sequences, given that it has~$j$ shotgun sequences in it:
\begin{equation}\label{eq:p11}
p_{1,1}(j)=\Bigl(1-\frac{\sigsize-1}{\sigsize}\fpooled_2\Bigr)^j-2p_{1,0}(j)-p_{0,0}(j).
\end{equation}
Let~$p_{1,+}(j)$ denote the event that the island 
consists of CAPS sequences from a fixed pool for one clone, 
at least one CAPS sequence for the other clone, and WGS sequences:
\begin{equation}\label{eq:p1x}
p_{1,+}(j)=\Bigl(1-\frac{\sigsize-1}{2\sigsize}\fpooled_2\Bigr)^j
	-p_{0,*}(j)-p_{1,0}(j).
\end{equation}
\end{subequations}
Using inclusion-exclusion,
\[
\Probcmd{\overline{D_k}}{E_2, M_k=j}
=\Bigl(2p_{0,*}(j)-p_{0,0}(j)\Bigr)
+2\sigsize p_{1,+}(j)-\sigsize^2 p_{1,1}(j).
\]
By Equations~(\ref{eq:p00}--\ref{eq:p1x}),
\begin{multline}\label{eq:map.fmap.j}
\Probcmd{\overline{D_k}}{E_2, M_k=j}
= 2\sigsize \Bigl(1-\frac{\sigsize-1}{2\sigsize}\fpooled_2\Bigr)^j
 - 2(\sigsize-1) \Bigl(1-\frac{\fpooled_2}2\Bigr)^j\\*
 - \sigsize^2 \Bigl(1-\frac{\sigsize-1}{\sigsize}\fpooled_2\Bigr)^j
 + 2\sigsize(\sigsize-1) \Bigl(1-\frac{\sigsize-\frac12}{\sigsize}\fpooled_2\Bigr)^j
 - (\sigsize-1)^2 (1-\fpooled_2)^j,
\end{multline}
which corresponds to Equation~\eqref{eq:pnd} 
with $\pnd(j)=\Probcmd{\overline{D_k}}{E_2, M_k=j}$.
Using the same technique as before
\[
1-p_2=\Probcmd{\overline{D_k}}{E_2}=\sum_{j=1}^{\infty}
\Probcmd{\overline{D_k}}{E_2, M_k=j}\Probcmd{M_k=j}{E_2},
\]
leading to Equation~\eqref{eq:prob.map}.

Recall that~$\pnd(j)$ is the probability of failing to map a
contig of~$j$ reads to the two clones simultaneously.
In order to show that the inequality in Equation~\eqref{eq:pnd}
holds, we prove that
\begin{equation}\label{eq:map.fmap.bound}
\pnd(j) < 2\sigsize \ndfactor^j-(2\sigsize-1)\ndfactor_3^j < 2\sigsize\ndfactor_1^j.
\end{equation}
Notice that $\ndfactor_5<\ndfactor_4<\ndfactor_3<\ndfactor_2<\ndfactor_1$ and thus  
$\pnd(j)\nearrow 2\sigsize\ndfactor_1^j$. Since~$\ndfactor_4=(\ndfactor_3+\ndfactor_5)/2$,
it follows from the convexity of~$x^j$ that
\begin{equation}\label{eq:b345}
2\ndfactor_4^j \le \ndfactor_3^j+\ndfactor_5^j.
\end{equation}
(Alternatively, notice that the same inequality follows from  
$p_{1,1}(j)\ge 0$ in Equation~\eqref{eq:p11}.)
We proceed by rearranging the equality of Equation~\eqref{eq:pnd}:
\begin{multline*}
2\sigsize \ndfactor_1^j - (2\sigsize-1) \ndfactor_3^j - \pnd(j)  = 
	2(\sigsize-1) \ndfactor_2^j
	+ (\sigsize-1)^2 \ndfactor_3^j
	- 2\sigsize(\sigsize-1) \ndfactor_4^j
	+ (\sigsize-1)^2 \ndfactor_5^j\\*
= 
	 (\sigsize-1)^2\underbrace{\Bigl(
		\ndfactor_3^j
		+\ndfactor_5^j
		-2\ndfactor_4^j
		\Bigr)}_{\text{$>0$ by Eq.~\eqref{eq:b345}}}
	+2(\sigsize-1) \underbrace{\Bigl(
		\ndfactor_2^j
		-\ndfactor_4^j
		\Bigr)}_{\text{$>0$ since $\ndfactor_2>\ndfactor_4$}},
\end{multline*}
which proves Equation~\eqref{eq:map.fmap.bound}.
\end{proof}

It is difficult to derive useful closed formulas
for the probability of overlap detection. 
For example, based on Equation~\eqref{eq:prob.map},
the number of contigs in the overlap that are simultaneously
mapped to the clones can be modeled as arrivals in a Poisson process 
with intensity $\coverage_2e^{-\coverage_2\roverlap}p_2$.
For practical values of~$\coverage_2$, this 
model seriously underestimates the probability of overlap detection.
The problem is similar to the one of using Lander-Waterman statistics \cite{LanderWaterman}
at high coverage levels (see \citeN{WendlWaterston} for a discussion).
For a more suitable model, let~$\rvgaps$ be the number of gaps entirely contained in the 
overlap, and number the islands from 0 to~$\rvgaps$. 
Let~$j_0, j_2, \dotsc, j_{\rvgaps}$ denote the number of shotgun sequences in
the islands. The probability that none of the islands can be 
mapped simultaneously to the two clones can be calculated as
\begin{equation}\label{eq:pnomap.exp}
p_{\mathrm{nomap}}(j_0,\dotsc,j_{\rvgaps})
	=\prod_{i=0}^{\rvgaps}\pnd(j_i),
\end{equation}
	where~$\pnd(j)$ is defined by Equation~\eqref{eq:pnd}.
(Notice that~$\rvgaps$ and the~$j_i$ are random variables.) 
We are interested in the expected value
$p_{\mathrm{nomap}}
	=\EXP p_{\mathrm{nomap}}(j_0,\dotsc,j_{\rvgaps})$.
In order to get a good assessment of CAPS-MAP performance, 
we found that
it is best to use a Monte-Carlo estimation\footnote{
Specifically, for every overlap size considered, 
we carried out a number of simulated experiments. 
Each experiment used a fixed number of 
shotgun sequences~$\rvreads$ placed 
randomly in the overlap, and
produced an instance of
a $(j_0,\dotsc,j_{\rvgaps})$ vector, for which 
$p_{\mathrm{nomap}}(j_0,\dotsc,j_{\rvgaps})$ 
was calculated using Equation~\eqref{eq:pnd}.
The average of these values was used to estimate $p_{\mathrm{nomap}}$.
The average was weighted with the probabilities 
of different~$\rvreads$ values, given by a Poisson distribution.
The set of~$\rvreads$ values was chosen so that 
it provided a sufficient 
accuracy for the weighted average estimate. 
For every~$\rvreads$, ten thousand experiments were 
performed.}
of this expected value; 
see Figure~\ref{fig:capsm.probs}. 
For an alternative, observe that
the inequality of
Equation~\eqref{eq:pnd} implies
$p_{\mathrm{nomap}}	< \EXP\Bigl[\ndfactor_1^{\rvreads}(2\sigsize)^{\rvgaps+1}\Bigr]$
where~$\rvreads$ is the number of sequences 
in the overlap, and thus $\rvreads=\sum_{i=0}^{\rvgaps} j_i$. 
Based on this observation,
we derived bounds (see Appendix) 
that are useful for large values of~$\coverage_2$ (e.g., $\coverage_2=7$),
but at lower coverages, this approach also 
underestimates the overlap detection probabilities significantly. 

\section{BAC ordering}
Our analyses so far have focused on detecting BAC overlaps via CAPS-MAP. 
This localized perspective was partly adopted to ease the theoretical analysis. 
In practice, mapping is performed based on a global clone-contig incidence matrix. 
The global approach exploits the dependencies in the collected data for increased 
accuracy.
The algorithmic issues are very similar  
to those encountered 
in the context of STS-based physical mapping
\shortcite{Gusfield}. 
Define the mapping matrix~$\mathbf{M}$ in which the rows correspond to the BACs,
the columns correspond to the contigs, and $\mathbf{M}[i,j]=1$
if contig~$j$ is linked to clone~$i$, otherwise $\mathbf{M}[i,j]=0$. 
We want to find the true ordering of the rows and columns, 
defined by their physical locations on the genome. 
Assume for a moment that the matrix is completely error-free,
i.e., all contigs are correctly assembled, and all contig-clone overlaps 
are detected.
It is not hard to see that the row and column permutations 
corresponding to the correct ordering result in a matrix~$\mathbf{M}'$ 
that satisfies the {\em consecutive ones} property (C1P) in the rows and the columns:
for every row~$i$, there exist $a\le b$ with $\mathbf{M}'[i,j]=1$ if and only if
$a\le j\le b$, and the same property holds for the columns. 
(A sufficient condition ensuring row-wise (or column-wise) C1P is that
	if the left endpoint of a contig (or a clone) precedes the left endpoint 
	of another one, the same holds for their right endpoints.)
Finding such permutations is a well-known problem~\shortcite{C1P},
and can be done in linear time.
When the matrix is not error-free, one can use 
techniques introduced for STS-based physical mapping. 
In \S\ref{sec:simulation} we detail a 
method that relies on traveling salesperson tours. 

\section{CAPS-MAP simulation of Drosophila assembly}\label{sec:simulation}
We tested the CAPS-MAP approach 
by simulating the assembly of the {\em Drosophila melanogaster}
genome. One of the main goals of the 
simulated assembly was to predict the performance a 
hybrid approach combining WGS
and CAPS sequences in a 
project that closely resembles the 
setup of the honey bee genome project's 
(\verb|http://www.hgsc.bcm.tmc.edu/projects/honeybee/|),
currently pursued at the Human Genome Sequencing Center
(HGSC)
of Baylor College of Medicine.

Concatenating all the Drosophila genome sequence (Release 2.5, 
112.6 million bases), 
2880 BAC sequences were generated 
by randomly picking their locations and lengths. The 
mean BAC insert length was 150 kbp, 
and its standard deviation was 500 bp.
The resulting random BAC library provides 3.6X coverage of the genome.
BACs were arrayed by first partitioning them into 5 groups, and then 
using a two-array $24\times24$ transversal design for each group. 
Every BAC was covered by 1.2X CAPS sequences: 0.4X per pool on 
the first array and 0.2X per pool on the second (reshuffled) array. 
In addition, WGS sequences were produced at 4X genome coverage. 
The shotgun sequences were generated using the program \texttt{wgs-simulator}
(written by K.~James Durbin),
which mimics shotgun sequence collection realistically by 
relying on sequence quality files \shortcite{Phred.error}
produced in sequencing projects.

Shotgun sequences were assembled into contigs using 
the Atlas suite of genome assembly tools (\verb|http://www.hgsc.bcm.tmc.edu/downloads/software/atlas/|) 
and Phrap (\verb|http://www.phrap.org/|).
A contig was mapped
to a clone if it contained sequences from all four clone pools.
Contigs that mapped to more than one 
BACs provided the evidence of BAC overlaps.
BACs were grouped into maximal overlapping sets, or {\em bactigs}.

We compared the overlap graphs to assess CAPS-MAP overlap detection.
The vertices of the overlap graphs are the BACs, and two BACs 
are connected if there is an overlap between them.
The true overlap graph for the original 
BACs contains 2880 vertices, and 10992 edges in 66 graph components. 
The overlap graph calculated from the bactigs 
has 9193 edges in 110 components. 
Among its edges, 8527 (93\%) are correct, 
and 2465 (22\%) of the true overlaps are not discovered. 
The median length of detected overlaps is 87 kbp, and the median length
of undetected overlaps is 42 kbp.
There are 666 edges that correspond to no real 
overlaps. The vast majority of these ``false positives''
are instances when a long read contig links several BACs,
which do not always overlap pairwise. 
All but two of the 
CAPS-MAP bactigs are true overlapping sets of BACs.
CAPS-MAP links the assembled contigs to BACs correctly even in these two 
bactigs:
the source of the error is the read contig assembly.
Table~\ref{tbl:droso} shows statistics on the bactig sizes and genome coverage.

\begin{table}
\begin{center}
\small
\begin{tabular}{|r|r|r|}
\hline
Minimum bactig size & Genome covered & Number of BACs in bactigs \\
\hline
2 & 97.1\% & 2758 \\
3 & 96.7\% & 2746 \\
5 & 94.9\% & 2714 \\
10 & 88.5\% & 2565 \\
15 & 82.4\% & 2400 \\
20 & 77.9\% & 2284 \\
30 & 65.0\% & 1945 \\ 
51 & 50.9\% & 1521 \\ 
60 & 40.3\% & 1195 \\
\hline
\end{tabular}
\end{center}
\caption{Statistics for simulated Drosophila assembly.
This table details the genome and BAC library coverage 
by bactig sizes. More than half of the genome is covered by bactigs 
with at least 51 BACs in them, defining the N50 
statistic for the clone map.
}\label{tbl:droso}
\end{table} 

BACs were ordered within each bactig.
For every bactig, an overlap matrix~$\mathbf{M}$ was
calculated, in which  
the rows correspond to the bactig's clones,
the columns correspond to the contigs linked to at least one bactig clone, 
and $\mathbf{M}[i,j]=1$
if contig~$j$ is linked to clone~$i$, otherwise $\mathbf{M}[i,j]=0$. 
The following
traveling salesperson (TSP)
formulation is used to find the correct column permutation. 
We search for a tour in a graph, in which every vertex corresponds to a 
contig (and thus a column), with an additional vertex~$u_0$. 
The weight of an edge between vertices~$u$ and~$u'$, corresponding 
to contigs~$j$ and~$j'$, is the 
number of rows in which they differ:
$w(u,u')=\sum_i\chi\Bigl\{\mathbf{M}[i,j]\ne\mathbf{M}[i,j']\Bigr\}$,
where~$\chi\{\cdot\}$ is the indicator function.
The weight of an edge between~$u$ and~$u_0$ is the sum of ones
in the column~$j$ that corresponds to~$u$: $w(u,u_0)=\sum_i\mathbf{M}[i,j]$.
Now, a Hamilton path with the minimum weight in this graph 
gives the best column permutation in the sense that it 
minimizes the number of gaps between blocks of ones within rows
\cite{mapping.tsp}.
The best row ordering could be found in an analogous manner, but we used
a simpler method which worked better in practice.
Clones are ordered relatively to the contig order 
by placing clone~$\clone$ before~$\clone'$ if
the first contig $\clone$ is linked to is before the first 
contig $\clone'$ is linked to, 
or if their first contigs are identical 
but $\clone$ has its last contig before $\clone'$.

\begin{figure}
\centerline{\includegraphics[height=.15\textheight]{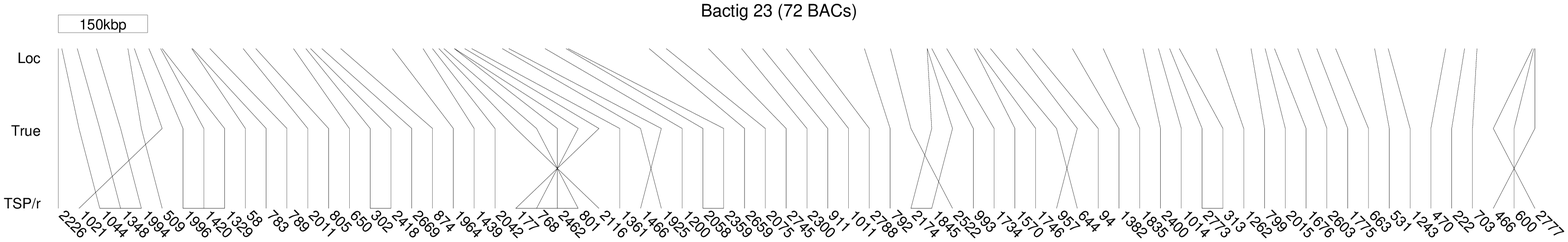}}
\centerline{\includegraphics[height=.15\textheight]{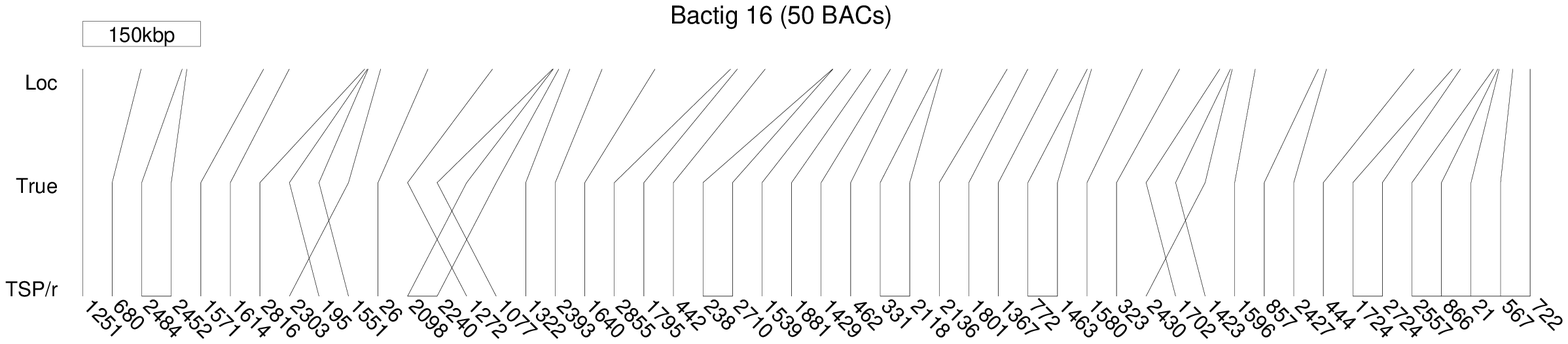}}
\caption{
Correctness of BAC ordering in Drosophila simulation.
The top (\textsf{Loc}) 
of each graph shows the relative 
physical location of each BAC,
the middle (\textsf{True}) shows 
the correct BAC order, and
the bottom (\textsf{TSP}) 
shows the TSP order,
and the BAC identifiers.
Identical BACs
are connected in order to
display the differences between the two permutations. 
The order of BACs at the bottom is not resolved when they are connected
with a horizontal line.
By resolving them optimally,
bactig 23 produces the order of 72 BACs with 12 breakpoints 
and bactig 16 orders 50 BACs with 9 breakpoints.
(Breakpoints are neighbors in 
the TSP order that are not neighbors in the true order.)
}
\label{fig:fly.order}
\end{figure}

We used the \texttt{concorde} program \shortcite{concorde} 
to solve the TSP instances.
The resulting row permutation is 
then further analyzed to find clones, for which the 
permutation arbitrarily enforces an order. Specifically, 
if consecutive rows of the permuted matrix~$\mathbf{M}'$ are identical,
then the order of the corresponding clones is not resolved.
Subsequently, we compared the TSP orders to the true orders, which 
is known since the BAC sequences are generated artificially.
Figure~\ref{fig:fly.order} shows the 
outcome of the comparison for two bactigs.
The TSP order is very close to the true order.

\section{Discussion}
The experimental expedience of shotgun sequencing has been essential 
for the success of genome-scale sequencing projects in the past decade. 
The power of the concept comes from the now established 
fact that the loss of information about read localization
incurred by random subcloning can be largely recovered 
in the assembly step using sequence information. 
Clone pooling is similar in spirit to shotgun sequencing in that it 
introduces experimental expedience by dramatically reducing the number 
of subclone library preparations. The clone pooling step leads to a 
temporary loss of information about localization of shotgun sequences on 
individual BAC clones. We have demonstrated that
sequence information can be used to 
successfully recover most of the information lost in
pooling.

Our analyses presented here indicate the theoretical feasibility of the CAPS-MAP
method and provide guidance for the design of genome-scale CAPS-MAP
experiments. In particular, our analysis indicates that transversal
pooling designs can accommodate high levels of clone redundancy and
perform well even at low levels of shotgun sequence coverage of clone
pools. 

Practical biological and technical considerations may set a limit to the
array size. In case of large genomes, the limitations may imply that the set of
BACs is partitioned and that pooling is applied separately to individual
subsets. This results in a lower clone redundancy within individual arrays
and a larger number of pools. Our analysis allows for the
partitioning of clones. It also allows for the
possibility of including whole-genome shotgun sequence reads. 
It thus covers 
realistic and practical scenarios of the CAPSS and CAPS-MAP methods' 
application.

\section*{Acknowledgements}
We are grateful to
Richard Gibbs and George
Weinstock for sharing pre-publication information on CAPSS and for useful
comments. 
Our discussion of 
computing CAPS-MAP overlap detection probabilities
has greatly benefited from conversations with 
Luc Devroye and Michael Waterman. 
This work was supported by grants
RO1~HG02583-01 from NHGRI at the NIH, 
U01~RR18464 from the NCRR,
and 
250391-02 from the NSERC.

\textsc{Remark.}\ \ 
An extended abstract of this paper is published in 
Genome Informatics vol.14 Universal Academy Press, Tokyo
(Proceedings of the  
14th International Conference on Genome Informatics (GIW),
December 14--17, 2003, Yokohama, Japan).

\bibliographystyle{chicago}

\clearpage
\appendix
\section*{Appendix}
Here we expand our discussion on the probability of overlap detection in CAPS-MAP.
In particular, we derive formulas that
show the exponential decay of the probability of not detecting an overlap
when the coverage~$\coverage_2$ is not too small.
We start with the bound
\begin{equation}\label{eq:pnomap.bound.def}
p_{\mathrm{nomap}}	< \EXP\Bigl[\ndfactor_1^{\rvreads}(2\sigsize)^{\rvgaps+1}\Bigr]
\end{equation}

Define
\[
\gengaps_{\nreads}(z) = \Expcmd{z^{\rvgaps}}{\rvreads},
\]	
the probability generating function for 
the distribution of the number of gaps conditioned on the number of shotgun sequences.
Define the events~$A_i$ for $i=1,\dotsc,\nreads-1$: $A_i$ denotes 
the event that the $i$-th sequence is followed by a gap, conditioned 
on the event $\{\rvreads=r\}$.
For arbitrary~$\ngaps$, and set of indexes $i_1<i_2<\dotsb<i_{\ngaps}$,
\[
\PROB\Bigl\{A_{i_1}A_{i_2}\dotsm A_{i_{\ngaps}}\Bigr\} 
	= (1-\ngaps\delta)_{+}^{\nreads},
\]
where $\delta=\frac{\roverlap\flength}{\coverlap\clength}$,
and $(x)_+=\max\{0,x\}$ \shortcite{EG,WendlWaterston}.
Let
\begin{align*}
S_0 & = 1\\*
S_g & = \sum_{i_1<\dotsb< i_{\ngaps}} \PROB\Bigl\{A_{i_1}A_{i_2}\dotsm A_{i_{\ngaps}}\Bigr\} 
	= \binom{\nreads-1}{\ngaps}(1-\ngaps\delta)_{+}^{\nreads}.
\end{align*}
Using inclusion-exclusion,
\[
\Probcmd{\rvgaps=\ngaps}{\rvreads=\nreads}
=\sum_{j=\ngaps}^{\nreads-1}
	\binom{j}{\ngaps}(-1)^{j-\ngaps} S_j.
\]
Hence,
\begin{align*}
\gengaps_{\nreads}(z) & = 
\sum_{\ngaps=0}^{\nreads-1}
z^\ngaps \sum_{j=\ngaps}^{\nreads-1} \binom{j}{\ngaps}(-1)^{j-\ngaps} S_j\\*
& = \sum_{j=0}^{\nreads-1} S_j
	\sum_{\ngaps=0}^j (-1)^{j-\ngaps} \binom{j}{\ngaps} z^{\ngaps} \\*
& = \sum_{j=0}^{\nreads-1} S_j (z-1)^j.
\end{align*}
Substituting the $S_j$ values:
\begin{equation}\label{eq:gengaps}
\gengaps_{\nreads}(z) =
	\sum_{j=0}^{\nreads-1} 
		\binom{\nreads-1}{j} (1-j\delta)_{+}^{\nreads} (z-1)^j,
\end{equation}
a result interesting on its own.

\begin{figure}
\centerline{\includegraphics[height=.3\textheight]{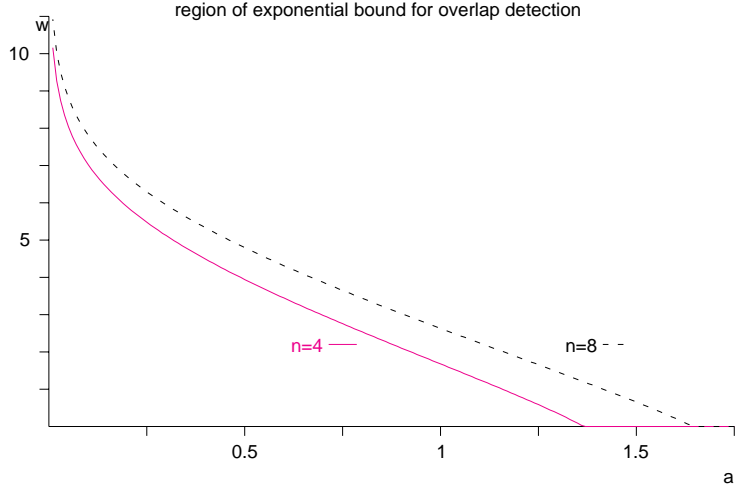}}
\caption{Values of the pooled shotgun coverage~$\pcoverage$
and WGS coverage~$\wgscoverage$, for which the clone overlap detection bound applies,
are above the graphs (see Equation~\eqref{eq:goodaw})}\label{fig:goodaw}. 
\end{figure}

Returning to Equation~\eqref{eq:pnomap.bound.def}, we have
\begin{equation}\label{eq:pnomap.bound.1}
p_{\mathrm{nomap}} 
< \EXP\biggl[
	2\sigsize \ndfactor_1^{\rvreads}
	 \sum_{j=0}^{\rvreads-1} 
		\binom{\rvreads-1}{j} (1-j\delta)_{+}^{\rvreads} (2\sigsize-1)^j
	\biggr],
\end{equation}
where~$\rvreads$ is a Poisson random variable with 
mean
\[
\mreads=
\frac{\coverage_2\coverlap\clength}{\flength}
\]
For every~$\nreads\ge0$, 
$(1-j\delta)_{+}^{\nreads} \le e^{-j\nreads\delta}$,
hence
\[
\sum_{j=0}^{\nreads-1} 
		\binom{\nreads-1}{j} (1-j\delta)_{+}^{\nreads} (2\sigsize-1)^j
	\le \Bigl(1+(2\sigsize-1)e^{-\nreads\delta}\Bigr)^{\nreads-1}.
\]
Consequently, by Equation~\eqref{eq:pnomap.bound.1},
\[
p_{\mathrm{nomap}} 
< \EXP\biggl[
	2\sigsize \ndfactor_1^{\rvreads}
		\Bigl(1+(2\sigsize-1)e^{-\rvreads\delta}\Bigr)^{\rvreads-1}
	\biggr].
\]
Recall that the random value we take the expectation of
is an upper bound on~$p_{\mathrm{nomap}}(j_0,\dotsc,j_{\rvgaps})$,
and thus if it is larger than one, it is useless.
Let
\[
f(\nreads)=
	\min\Bigl\{1,2\sigsize \ndfactor_1^{\nreads}
		\Bigl(1+(2\sigsize-1)e^{-\nreads\delta}\Bigr)^{\nreads-1}\Bigr\}.
\]
So we have in fact the bound
\begin{equation}\label{eq:pnomap.bound.2}
p_{\mathrm{nomap}}
	< \EXP f(\rvreads).
\end{equation}
In order to achieve exponential decay in the bound, we would like to have
\[
\ndfactor_1\Bigl(1+(2\sigsize-1)e^{-\nreads_0\delta}\Bigr)<1
\]
for some~$\nreads_0<\mreads$. Rearranging the inequality, we have
\begin{equation}\label{eq:goodaw}
(2\sigsize-1)\frac{\sigsize(\pcoverage+\wgscoverage)+\pcoverage}{(\sigsize-1)\pcoverage}
	< e^{(2\pcoverage+\wgscoverage)\roverlap},
\end{equation}
which is satisfied when~$\pcoverage$ and~$\wgscoverage$ are not too small 
(see Figure~\ref{fig:goodaw}).

There are several possible ways to 
exploit the fact that the exponential component of $f(\nreads)$ becomes
for~$\nreads$ less than the expected value~$\mreads$.
The main idea is that when evaluating
$\EXP f(\rvreads)=\sum f(\nreads)\PROB\{\rvreads=\nreads\}$
in Equation~\eqref{eq:pnomap.bound.2}, 
either the probability of~$\rvreads=\nreads$ is small, or
the value of~$f(\nreads)$ is small.
Let~$0<k<\lambda$ be a threshold (that we specify later), and let~$\alpha=k/\mreads$.
To proceed with Equation~\eqref{eq:pnomap.bound.2}, we condition
on the event~$\{\rvreads\le\alpha\mreads\}$.
We use the bound 
\begin{equation}\label{eq:poisson.bound}
\PROB\{\rvreads\le \alpha\mreads\}
	< \frac{e^{-\mreads(1-\alpha)^2/2}}{(1-\alpha)\sqrt{2\pi\alpha\mreads}},
\end{equation}
which we prove here quickly.
By definition,
\begin{align*}
\PROB\{\rvreads\le\alpha\mreads\}
& \le \sum_{\nreads=0}^k
	\frac{{\mreads}^{\nreads}}{\nreads!} e^{-\mreads}	
< e^{-\mreads}\frac{{\mreads}^{k}}{k!}
	\sum_{\nreads=0}^k \Bigl(\frac{k}{\mreads}\Bigr)^{\nreads}\\*
& <  e^{-\mreads}\frac{{\mreads}^{k}}{k!} (1-\alpha)^{-1} 
< e^{-\mreads(1-\alpha+\alpha\ln\alpha)} \frac{1}{(1-\alpha)\sqrt{2\pi\alpha\lambda}},
\end{align*}
where we used a Stirling approximation: $k!>(k/e)^k/\sqrt{2\pi k}$. Using a Taylor series expansion, 
\[
1-\alpha+\alpha\ln\alpha = \frac12 (1-\alpha)^2 + \frac16 (1-\alpha)^3 + \frac{1}{12}(1-\alpha)^4 \dotsc
\]
and thus $1-\alpha+\alpha\ln\alpha>\frac12 (1-\alpha)^2$ for~$0<\alpha<1$,
and Equation~\eqref{eq:poisson.bound} follows.

\begin{figure}
\centerline{\includegraphics[width=\textwidth]{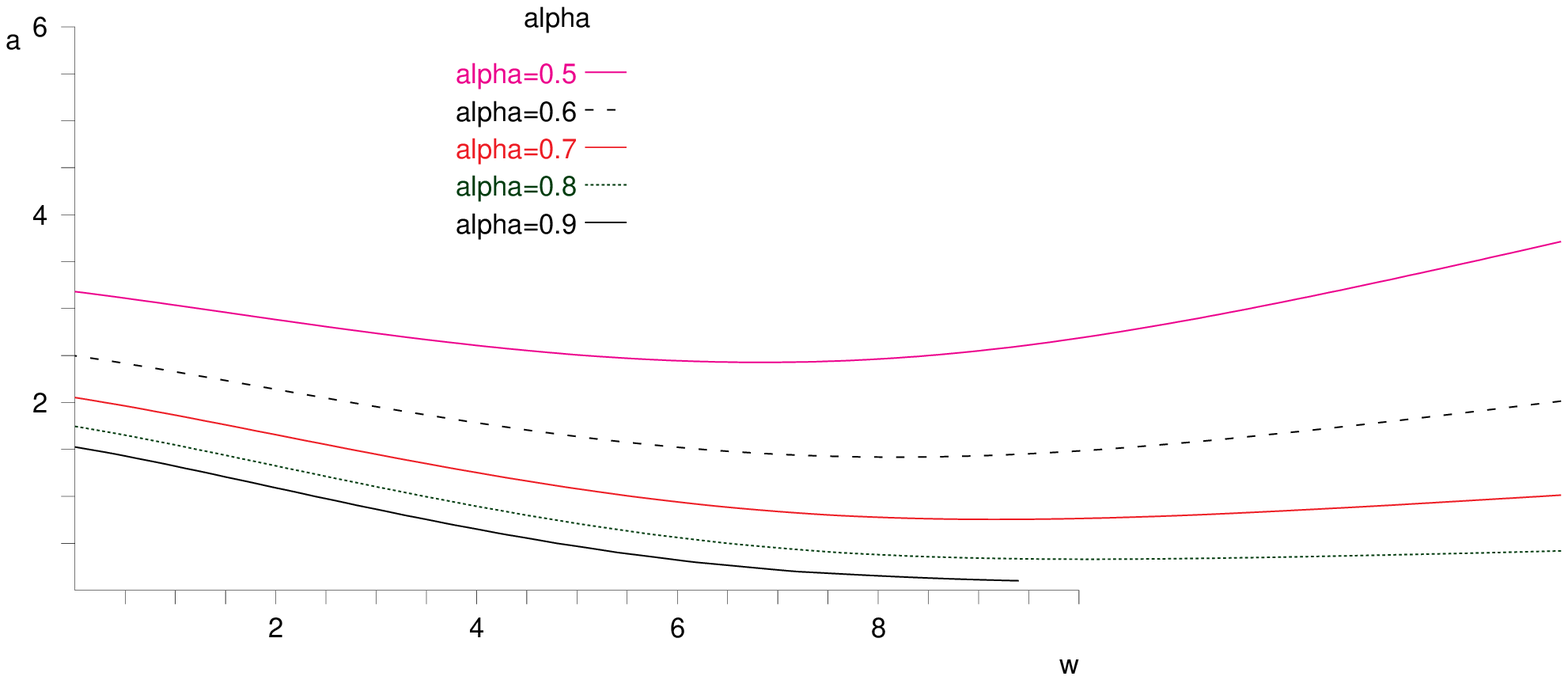}}
\caption{Balanced $\alpha$ values for our exponential bound.}\label{fig:balance.alpha}
\end{figure}

Now,
\begin{align*}
\EXP f(\rvreads)
 & = \Expcmd{f(\rvreads)}{\rvreads\le\alpha\mreads}\PROB\{\rvreads\le\alpha\mreads\}
 +\Expcmd{f(\rvreads)}{\rvreads>\alpha\mreads} \PROB\{\rvreads>\alpha\mreads\}\\*
 & \le \PROB\{\rvreads\le\alpha\mreads\} + \Expcmd{f(\rvreads)}{\rvreads>\alpha\mreads}\\*
 & < \frac{e^{-\mreads(1-\alpha)^2/2}}{(1-\alpha)\sqrt{2\pi\alpha\mreads}}
 	+ \frac{2\sigsize e^{-\mreads}
 		\sum_{\nreads=0}^{\infty}
 			\frac{\Bigl(\ndfactor_1(1+(2\sigsize-1)e^{-\alpha\delta\mreads})\Bigr)^{\nreads}}{\nreads!}}{1+(2\sigsize-1)e^{-\alpha\delta\mreads}}
 		\\*
 & = \frac{\exp\Bigl(-\mreads(1-\alpha)^2/2\Bigr)}{(1-\alpha)\sqrt{2\pi\alpha\mreads}}
 	+ \frac{2\sigsize \exp\biggl(-\mreads\Bigl(1-\ndfactor_1(1+(2\sigsize-1)e^{-\alpha\coverage_2\roverlap})\Bigr)\biggr)}{
 		1+(2\sigsize-1)e^{-\alpha\coverage_2\roverlap}},
 \end{align*}
 where we used~$\delta\mreads=\coverage_2\roverlap$.
Figure~\ref{fig:balance.alpha} shows values of~$\alpha$ for different~$\pcoverage,\wgscoverage$ pairs 
that balance the exponents in the two terms. 

After choosing a balancing~$\alpha$ value for a given~$(\pcoverage,\wgscoverage)$ pair, 
we obtain
\[ 
\EXP f(\rvreads) < X_1 \exp(-X_2 \roverlap\clength),
\]
where~$X_1$ and~$X_2$ are constants that do not depend on~$\roverlap$.
The bound becomes small ($<10^{-8}$)
for larger~$\coverage_2$ values (e.g., $\coverage_2=7$),
but even then, it is not very tight. 
Based on simulation results, the tightness is lost 
with the inequality of Equation~\eqref{eq:pnomap.bound.def},
and not in the following steps.
For example, we evaluated the bounds of Equations~\eqref{eq:pnomap.bound.1} 
and~\eqref{eq:pnomap.bound.2} numerically.
While they are fairly close to each other, and to the exponential bound 
using~$\alpha$, they already bound the expected value of~Equation~\eqref{eq:pnomap.exp} 
rather loosely in many cases.
Furthermore, even for~$(\pcoverage,\wgscoverage)$ pairs
for which we cannot establish exponential decay 
using the inequality of Equation~\eqref{eq:pnomap.bound.def}, the overlap
detection probability may get very close to one. 
For instance, a two-array design with
$\pcoverage=0.5$ and~$\wgscoverage=2$ falls below the curve
of Figure~\ref{fig:goodaw}, yet can be employed efficiently 
in CAPS-MAP as shown in Figure~\ref{fig:capsm.probs}.
Therefore, we prefer using a Monte-Carlo evaluation of Equation~\eqref{eq:pnomap.exp}
to predict the experimental performance of CAPS-MAP.\label{veryend}

\end{document}